\documentclass[aps,twocolumn,pra,preprintnumbers,showpacs,tightenlines]{revtex4}
\usepackage{amssymb}
\usepackage{amsmath}
\usepackage{graphicx}
\usepackage{epsfig}
\usepackage{subfigure}
\usepackage{amsfonts}
\usepackage{CJK}

\begin{document}

\title{Efficient transfer of an arbitrary qutrit state in circuit QED}

\author{Tong Liu, Shao-Jie Xiong, Xiao-Zhi Cao, Qi-Ping Su,}
\author{Chui-Ping Yang}\email{yangcp@hznu.edu.cn}
\address{Department of Physics, Hangzhou Normal University, Hangzhou, Zhejiang 310036, China }

\begin{abstract}
Compared with a qubit, a qutrit (i.e., three-level quantum system) has a larger Hilbert space and thus can be used to encode more information in quantum information processing and communication. Here, we propose a method to transfer an arbitrary quantum state between two flux qutrits coupled to two resonators. This scheme is simple because it only requires two basic operations. The state-transfer operation can be performed fast because of using resonant interactions only. Numerical simulations show that high-fidelity transfer of quantum states between the two qutrits is feasible with current circuit-QED technology. This scheme is quite general and can be applied to accomplish the same task for other solid-state qutrits coupled to resonators.
\end{abstract}

 \maketitle
\date{\today}

Superconducting qubits for quantum information and quantum computation have attracted considerable attention due to their controllability, ready fabrication,
integrability, and potential scalability [1-5]. Their coherence time has recently been significantly increased [6-11]. For superconducting qubits, the level spacings can be rapidly adjusted within 1-3 ns, by varying external control parameters (e.g., magnetic flux applied to the superconducting loop of a superconducting phase, transmon, Xmon or flux qubit; see, e.g.,~[8,12-14]). In addition, circuit QED, i.e., analogs of cavity QED with solid-state systems, has been considered as one of the most promising candidates for building quantum computers and quantum information processors [3-5,15,16]. Furthermore, the strong-coupling or ultrastrong-coupling regime of a qubit with a microwave cavity has been reported in experiments~[17,18].

Quantum information processing (QIP) with qudits ($d$-level systems), including qutrits, has been attracting increasing interest. For example, QIP and tomography of nanoscale semiconductor devices were studied (see, Refs.~[19,20] and references therein). On the other hand, quantum state transfer (QST) plays an important role in quantum communication and QIP. Over the past years, theoretical proposals have been presented for realizing qubit-to-qubit QST with two superconducting qubits, which are coupled through a cavity/resonator~[21-28] or a capacitor [29]. The cavity/resonator acts as a quantum data bus to mediate long-range and fast interaction between superconducting qubits. The QST between two superconducting qubits has been demonstrated in circuits consisting of superconducting qubits coupled to cavities or resonators [30-33]. However, to the best of our knowledge, there is no study of qutrit-to-qutrit QST in circuit QED.

\begin{figure}[tbp]
\begin{center}
\includegraphics[bb=3 3 584 411, width=6.5 cm, clip]{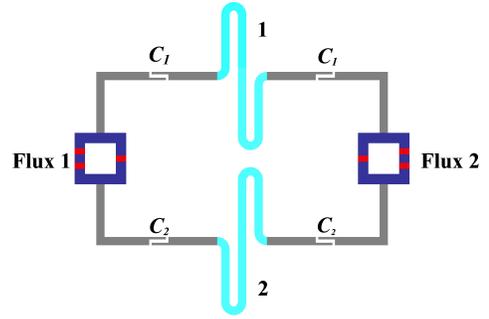} %
\vspace*{-0.08in}
\end{center}
\caption{Diagram of a setup for two flux qutrits coupled to two TLRs via capacitances $C_1$ and $C_2$. A blue
square represents a flux qutrit, which can be other types of solid-state qutrit, such as a quantum dot, a superconducting phase qutrit or a transmon qutrit.}
\label{fig:1}
\end{figure}

In this letter, we propose a scheme to transfer an arbitrary quantum state between two superconducting flux qutrits, coupled to two transmission line resonators~(TLRs)~(Fig.~1). The three levels of qutrit $j$ are denoted as $|g\rangle_j,$ $|e\rangle_j,$ and $|f\rangle_j$ $(j=1,2)$ (Fig.~2). The QST from qutrit 1 to qutrit 2 is expressed by the formula
\begin{eqnarray}
(\alpha|g\rangle_1+\beta|e\rangle_1+\gamma|f\rangle_1)|g\rangle_2\nonumber \\
\rightarrow|g\rangle_1(\alpha|g\rangle_2+\beta|e\rangle_2+\gamma|f\rangle_2),
\end{eqnarray}
 where $\alpha,$ $\beta,$ and $\gamma$ are the normalized complex numbers; the subscripts 1 and 2 represent qutrit 1 and qutrit 2.

This scheme has the following advantages.~Firstly, it is simple because the state transfer requires only two basic operations. Secondly, the speed of operation is fast because of using resonant interactions, e.g., the QST can be completed within $\sim 8.5$ ns, as shown in our following numerical simulation. Thirdly, through the numerical simulation, we find that precise control of qutrit-resonator coupling and qutrit-resonator resonance is not necessary and the qutrit-to-qutrit QST with a high fidelity is feasible with current circuit-QED technology. Lastly, this scheme is quite general and can be applied to implement QST between other solid-state qutrits, e.g., other types of superconducting qutrits or quantum dots coupled to resonators. We hope this work will stimulate experimental activities in the near future.
\begin{figure}[tbp]
\begin{center}
\includegraphics[bb=16 422 573 666, width=8.2 cm, clip]{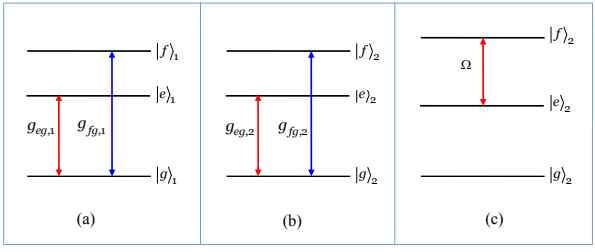}
\vspace*{-0.08in}
\end{center}
\caption{(a) TLR 1 (2) is resonant with the
$|g\rangle\leftrightarrow|e\rangle $ ($|g\rangle\leftrightarrow|f\rangle $) transition of qutrit 1 with a coupling constant $g_{eg,_1}$ ($g_{fg,_1}$).
(b) TLR 1 (2) is resonant with $|g\rangle\leftrightarrow|e\rangle$ ($|g\rangle\leftrightarrow|f\rangle $) transition of qutrit 2 with a coupling constant $g_{eg,_2}$ ($g_{fg,_2}$). (c) The classical pulse is resonant with the $|e\rangle\leftrightarrow|f\rangle $ transition of qutrit 2, with a Rabi frequency $\Omega$.}
\label{fig:2}
\end{figure}

Consider a system of two flux qutrits connected by two TLRs (Fig.~1). As shown in Fig.~2(a,b), TLR~1 is resonantly coupled to the
$|g\rangle\leftrightarrow|e\rangle $ transition of qutrit $j$ with a coupling constant $g_{eg,_j}$, TLR~2 is resonantly coupled to the $|g\rangle\leftrightarrow|f\rangle $ transition of qutrit $j$ with a coupling constant $g_{fg,_j}$ $(j=1,2)$. In the interaction picture, the Hamiltonian is given by (in units of $\hbar =1$)
\begin{eqnarray}
H_{I_,1} &=& \sum_{j=1}^2g_{eg,_j}(a_1\sigma_{eg,_j}^{+}+h.c.) \nonumber \\
&+& \sum_{j=1}^2g_{fg,_j}(a_2\sigma_{fg,_j}^{+}+h.c.),
\end{eqnarray}
where $a_j$ is the photon annihilation operator for the TLR $j$, $\sigma_{eg,_j}^{+}=|e\rangle _{j}\langle g|$ and $\sigma_{fg,_j}^{+}=|f\rangle _{j}\langle g|$ $(j=1,2)$. For simplicity, we set $g_{eg,j}=g_{fg,_j}\equiv g$, which can be achieved by a prior design of the sample with appropriate capacitances $C_1$ and $C_2$.

Under the Hamiltonian~(2), one can obtain the following state evolutions
\begin{eqnarray*}
|g\rangle|0\rangle|0\rangle|g\rangle&\rightarrow&|g\rangle|0\rangle|0\rangle|g\rangle, \nonumber\\
|e\rangle|0\rangle|0\rangle|g\rangle&\rightarrow&\frac{1}{2}(1+\cos\sqrt{2}gt)|e\rangle|0\rangle|0\rangle|g\rangle\nonumber\\
&-&\frac{\sqrt{2}}{2}i\sin(\sqrt{2}gt)|g\rangle|1\rangle|0\rangle|g\rangle\nonumber \\
\end{eqnarray*}
\begin{eqnarray}
&-&\frac{1}{2}(1-\cos\sqrt{2}gt)|g\rangle|0\rangle|0\rangle|e\rangle,\nonumber \\
|f\rangle|0\rangle|0\rangle|g\rangle&\rightarrow&\frac{1}{2}(1+\cos\sqrt{2}gt)|f\rangle|0\rangle|0\rangle|g\rangle\nonumber\\
&-&\frac{\sqrt{2}}{2}i\sin(\sqrt{2}gt)
|g\rangle|0\rangle|1\rangle|g\rangle\nonumber \\
&-&\frac{1}{2}(1-\cos\sqrt{2}gt)|g\rangle|0\rangle|0\rangle|f\rangle.
\end{eqnarray}
Here and below, the left $|0\rangle$ ($|1\rangle$) in each term represents the vacuum state (the single photon state) of TLR 1 while
the right $|0\rangle$ ($|1\rangle$) in each term represents the vacuum state (the single photon state) of TLR 2; in addition, the left $|g\rangle$, $|e\rangle$, and $|f\rangle$ in each term are the states of qutrit 1 while the right $|g\rangle$, $|e\rangle$, and $|f\rangle$ in each term are the states of qutrit 2.

From Eq.~(3), it can be seen that when the interaction time is equal to $t_1=\pi/(\sqrt{2}g)$, one obtains the transformations
$|g\rangle|0\rangle|0\rangle|g\rangle\rightarrow|g\rangle|0\rangle|0\rangle|g\rangle$, $|e\rangle|0\rangle|0\rangle|g\rangle\rightarrow-|g\rangle|0\rangle|0\rangle|e\rangle$,
$|f\rangle|0\rangle|0\rangle|g\rangle\rightarrow-|g\rangle|0\rangle|0\rangle|f\rangle$ simultaneously. As a result, we have the
following state transformation
\begin{eqnarray}
(\alpha|g\rangle+\beta|e\rangle+\gamma|f\rangle)|0\rangle|0\rangle|g\rangle\nonumber \\
\rightarrow|g\rangle|0\rangle|0\rangle(\alpha|g\rangle-\beta|e\rangle-\gamma|f\rangle).
\end{eqnarray}

Adjust the level spacings of each qutrit so that it is decoupled from the two TLRs. Then apply a classical pulse to qutrit 2. The pulse is resonant with the $|e\rangle\leftrightarrow|f\rangle$ transition of qutrit 2~[Fig.~2(c)].
The Hamiltonian in the interaction picture is expressed as
\begin{eqnarray}
H_{I_,2}=\Omega(|e\rangle _{2}\langle
f|+h.c.),
\end{eqnarray}
where $\Omega$ is the Rabi frequency of the pulse. One can obtain the following rotations under the
Hamiltonian~(5),
\begin{eqnarray}
|e\rangle _{2} &\rightarrow &\cos (\Omega t)|e\rangle _{2}-i\sin (\Omega t)|f\rangle _{2},  \nonumber \\
|f\rangle _{2} &\rightarrow &\cos (\Omega t)|f\rangle _{2}-i\sin (\Omega t)|e\rangle _{2}.
\end{eqnarray}
We set $t_2=\pi/\Omega$ to obtain a $\pi$ rotation by $|e\rangle _{2}\rightarrow-|e\rangle _{2}$ and $|f\rangle _{2}\rightarrow-|f\rangle _{2}$. Hence, we can obtain $-|g\rangle|0\rangle|0\rangle|e\rangle\rightarrow|g\rangle|0\rangle|0\rangle|e\rangle$ and $-|g\rangle|0\rangle|0\rangle|f\rangle\rightarrow|g\rangle|0\rangle|0\rangle|f\rangle$. Therefore, it follows from Eq.~(4)
\begin{eqnarray}
(\alpha|g\rangle+\beta|e\rangle+\gamma|f\rangle)|0\rangle|0\rangle|g\rangle\nonumber \\
\rightarrow|g\rangle|0\rangle|0\rangle(\alpha|g\rangle+\beta|e\rangle+\gamma|f\rangle),
\end{eqnarray}
which shows that the original state of qutrit 1 is perfectly transferred to qutrit 2 after the above operation.

Let us now give a discussion of the experimental implementation. For a general consideration, we will analyze the fidelity of the QST by allowing a small qutrit-resonator frequency detuning. In addition, we take into account the crosstalk between the two resonators. Thus, the Hamiltonian~(2) is modified as follows
\begin{small}
\begin{eqnarray}
\widetilde{H}_{I,1}&=&\sum_{j=1}^2g_{eg,_j}(e^{i\delta
_1t}a_1\sigma_{eg,_j}^{+}+h.c.)\nonumber \\
&+&\sum_{j=1}^2g_{fg,_j}(e^{i\delta
_2t}a_2\sigma_{fg,_j}^{+}+h.c.)\nonumber \\
&+&g_{12}\left(  e^{i\Delta t}a_1a_2^{\dagger}+h.c.\right).
\end{eqnarray}
\end{small}
\hspace{-0.05in}The first (second) term of Eq.~(8)
describes the coupling between TLR~1 (2) and the $\left| g\right\rangle \leftrightarrow \left|
e\right\rangle$ ($\left| g\right\rangle \leftrightarrow \left|
f\right\rangle)$ transition of qutrit $j$, with detuning $\delta_1=\omega _{eg_j}-\omega
_{c_1}$ ($\delta_2=\omega _{fg_j}-\omega
_{c_2}$). Here, $\omega _{eg_j}$ ($\omega _{fg_j}$) is the $\left| g\right\rangle \leftrightarrow \left|
e\right\rangle$ ($\left| g\right\rangle \leftrightarrow \left|
f\right\rangle)$ transition frequency of qutrit $j$, and $\omega _{c_j}$ is the frequency of TLR $j$ $(j=1,2)$. For simplicity, we consider identical flux qutrits and set $\delta_1=\delta_2=\delta$. The last term of Eq.~(8) describes the inter-resonator crosstalk
between the two TLRs, where $\Delta =\omega _{c_2}-\omega_{c_1}$ is detuning between the two-TLR frequencies and $g_{12}$ is the inter-resonator crosstalk coupling strength.

We also consider the inter-resonator crosstalk coupling during the qutrit-pulse resonant interaction. Thus, the Hamiltonian~(5) is modified as
\begin{eqnarray}
\widetilde{H}_{I,2}=H_{I,2}+g_{12}\left(  e^{i\Delta t}a_1a_2^{\dagger}+h.c.\right).
\end{eqnarray}

The dynamics of the lossy system, with finite qutrit relaxation and
dephasing and photon lifetime included, is determined by the following
master equation
\begin{small}
\begin{eqnarray}
\frac{d\rho }{dt} &=&-i[ \widetilde{H}_{I},\rho ] +\sum_{j=1}^{2}\kappa _{j}
\mathcal{L}[ a_{j}]  \nonumber \\
&+&\sum_{j=1,2}\left\{ \gamma _{eg_j}\mathcal{L}[ \sigma
_{eg,_j}^{-}] +\gamma _{fe_j}\mathcal{L}[ \sigma _{fe,_j}^{-}]
+\gamma _{fg_j}\mathcal{L}[ \sigma _{fg,_j}^{-}] \right\}  \nonumber\\
&+&\sum_{j=1,2}\left\{ \gamma _{j,\varphi f}\left( \sigma _{ff_j}\rho
\sigma _{ff_j}-\sigma _{ff_j}\rho /2-\rho \sigma _{ff_j}/2\right) \right\} \nonumber \\
&+&\sum_{j=1,2}\left\{ \gamma _{j,\varphi e}\left( \sigma _{ee_j}\rho
\sigma _{ee_j}-\sigma _{ee_j}\rho /2-\rho \sigma _{ee_j}/2\right) \right\},
\end{eqnarray}
\end{small}
\hspace{-0.05in}where $\widetilde{H}_{I}$ is $\widetilde{H}_{I,1}$ or $\widetilde{H}_{I,2}$ above, $\sigma _{eg,_j}^{-}=\left\vert g\right\rangle _{j}\left\langle
e\right\vert$, $\sigma _{fe,_j}^{-}=\left\vert e\right\rangle _{j}\left\langle
f\right\vert$, $\sigma _{fg,_j}^{-}=\left\vert g\right\rangle _{j}\left\langle
f\right\vert , \sigma _{ee_j}=\left\vert e\right\rangle _{j}\left\langle
e\right\vert ,\sigma _{ff_j}=\left\vert f\right\rangle _{j}\left\langle
f\right\vert ;$ and $\mathcal{L}\left[ \Lambda \right] =\Lambda \rho \Lambda
^{+}-\Lambda ^{+}\Lambda \rho /2-\rho \Lambda ^{+}\Lambda /2,$ with $\Lambda
=a_{j},\sigma _{eg,_j}^{-},\sigma _{fe,_j}^{-},\sigma _{fg,_j}^{-}.$ Here, $\kappa
_{j}$ is the photon decay rate of TLR $j$. In addition, $
\gamma _{eg_j}$ is the energy relaxation rate of the level $\left\vert
e\right\rangle $ of qutrit $j$, $\gamma _{fe_j}$ ($\gamma _{fg_j}$) is the
energy relaxation rate of the level $\left\vert f\right\rangle $ of qutrit $
j $ for the decay path $\left\vert f\right\rangle \rightarrow \left\vert
e\right\rangle $ ($\left\vert g\right\rangle $), and $\gamma _{j,\varphi e}$
($\gamma _{j,\varphi f}$) is the dephasing rate of the level $\left\vert
e\right\rangle $ ($\left\vert f\right\rangle $) of qutrit $j$ ($j=1,2$).

The fidelity of the operation is given by
$\mathcal{F}=\sqrt{\left\langle \psi _{\mathrm{id}}\right\vert \rho
\left\vert \psi _{\mathrm{id}}\right\rangle},$
where $\left\vert \psi _{\mathrm{id}}\right\rangle $ is the output state $|g\rangle|0\rangle|0\rangle(\alpha|g\rangle+\beta|e\rangle+\gamma|f\rangle)$ of an ideal system (i.e., without dissipation, dephasing, and
crosstalk); while $\rho$ is the final density operator of the system when the operation is performed in
a realistic situation. As an example, we will consider the case of  $\alpha=\beta=\gamma=1/\sqrt{3}$ in the following analysis.

We now numerically calculate the fidelity of operation. We set $\Omega/2\pi=$100 MHz, which can be achieved in experiments~[34].
For a flux qutrit, the transition frequency between two neighbor levels is 1 to 30 GHz. Thus, we can choose $\omega_{c_1}/2\pi=4.5$ GHz and $\omega _{c_2}/2\pi=7.0$ GHz. Accordingly, we have $\Delta/2\pi=2.5$ GHz. We assume that the frequencies of the TLRs are fixed during the entire operation. Other parameters used in the numerical simulation are: (i) $\gamma^{-1} _{j,\varphi e}=1.5~\mu s,$ $\gamma ^{-1}_{j,\varphi f}=0.5~\mu s$; (ii) $\gamma^{-1} _{eg_j}=\gamma ^{-1}_{fe_j}=\gamma ^{-1}_{fg_j}=2.5~\mu s$; (iii) $\kappa^{-1} _{j}=20~\mu s$ $(j=1,2)$. We here consider a conservative case for the decoherence times of flux qutrits~[11].

\begin{figure}[tbp]
\begin{center}
\includegraphics[bb=6 208 493 496, width=8.0 cm, clip]{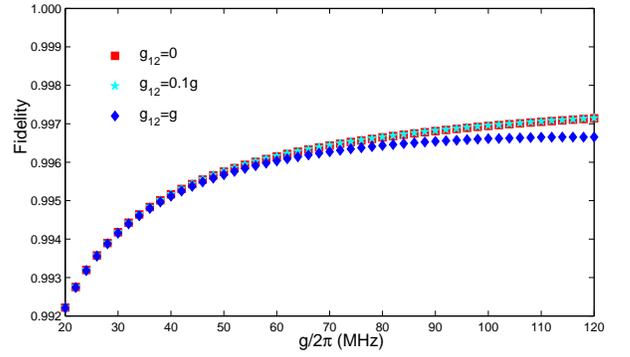}
\vspace*{-0.08in}
\end{center}
\caption{Fidelity versus $g/2\pi$, plotted  for $g_{eg,j}=g_{fg,_j}\equiv g$ ($j=1,2$); $g_{12}=0, 0.1g, g$; and $\delta=0$ (i.e., the qutrit-resonator resonance case). Here and in Fig.~4, $g_{12}$ is the inter-resonator crosstalk coupling strength and $\delta$ is the detuning.}
\label{fig:3}
\end{figure}

To start with, let us first investigate the effect of the inter-resonator crosstalk on the operational fidelity.~Figure~3 depicts the fidelity versus $g/2\pi$ with $g_{12}=0, 0.1g, g$, which is plotted for the qutrit-resonator resonance case.~From Fig.~3, one can see that the effect of the inter-resonator crosstalk is very small when $g_{12}\leq 0.1g$ and a high fidelity $\sim99.7\%$ can be reached for $g/2\pi=100$ MHz~[18]. In this case, the estimated operation time is $\sim 8.5$ ns. In the following, we choose $g_{12}=0.1g$, which is readily satisfied in experiments ~[35].

\begin{figure}[tbp]
\begin{center}
\includegraphics[bb=0 0 568 423, width=8.0 cm, clip]{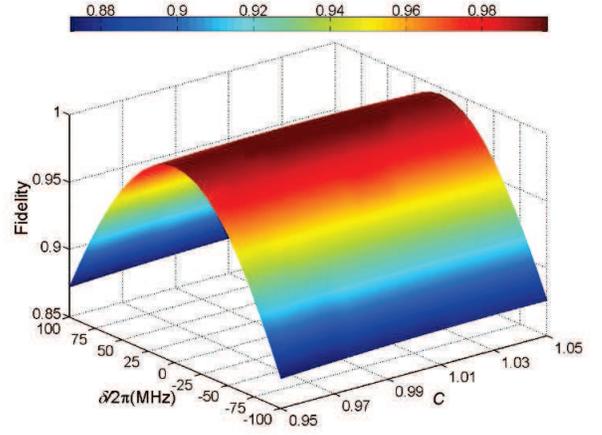}
\vspace*{-0.08in}
\end{center}
\caption{Fidelity versus $\delta/2\pi$ and $c$, plotted for $g_{12}=0.1g$ and $g/2\pi=100$ MHz. Here, $c=g_{fg,_j}/g_{eg,j}$, with $g_{eg,j}\equiv g$ ($j=1,2$).}
\label{fig:4}
\end{figure}

In realistic situation, it may be a challenge to obtain exact qutrit-resonator resonant interaction and identical qutrit-resonator coupling. Thus, we consider a small qutrit-resonator frequency detuning and inhomogeneous qutrit-resonator coupling. We set $g_{eg,1}=g_{eg,2}=g$ but $g_{fg,_1}=g_{fg,_2}=cg$, with $c\in[0.95,1.05]$. This may apply when the two qutrits $1$ and $2$, the two coupling capacitances $C_1$ (Fig.1), and the two coupling capacitances $C_2$ (Fig.~1) are approximately identical. Figure~4 shows the fidelity versus $\delta/2\pi$ and $c$, which is plotted for $g/2\pi=100$ MHz. From Fig.~4, one can obtain $\{{\mathcal{F}},\delta/2\pi,c\}$: (i) $\{$0.990, 20 MHz, 0.96$\}$; (ii) $\{$0.975, 40 MHz, 0.99$\}$; (iii) $\{$0.950, 60 MHz, 1.02$\}$; and (iv) $\{$0.915, 80 MHz, 1.05$\}$. Figure~4 shows that for $\delta/2\pi$ $\in[-80,80]$ MHz and $c\in[0.95,1.05]$, the fidelity is greater than $91\%$.

For TLRs 1 and 2 with frequencies and dissipation times used in the numerical simulation, the quality factors of the two TLRs are $Q_1=5.7\times10^5$ and $Q_2=8.8\times10^5$. The coplanar waveguide resonators with a loaded quality factor $Q\sim10^6$ have been implemented in experiments~[36,37].
We have numerically simulated a QST between two flux qutrits, which shows that the high-fidelity implementation of a qutrit-to-qutrit QST is feasible with current circuit-QED technology.

C. P. Yang was supported in part by the National Natural Science Foundation
of China under Grant Nos. 11074062 and 11374083, the Zhejiang Natural
Science Foundation under Grant No.~LZ13A040002, and the funds from Hangzhou
Normal University under Grant Nos.~HSQK0081 and PD13002004. This work was
also supported by the funds from Hangzhou City for the Hangzhou-City Quantum
Information and Quantum Optics Innovation Research Team.

\end{document}